\documentclass[preprint,tightenlines,aps,prd,showpacs,nofootinbib]{revtex4}
\usepackage{graphicx}
\usepackage{amsmath}
\usepackage{amssymb}
\usepackage{bm}
\usepackage{bbm}

\begin{document}

\title{\mbox{}\\[10pt]
$\Upsilon$ decay to two-charm quark jets  as a Probe of the Color
Octet Mechanism}


\author{Yu-Jie Zhang$~^{(a)}$
  and Kuang-Ta Chao$~^{(a,b)}$}
\affiliation{ {\footnotesize (a)~Department of Physics and State Key
Laboratory of Nuclear Physics and Technology, Peking University,
 Beijing 100871, China}\\
{\footnotesize (b)~Center for High Energy Physics, Peking
University, Beijing 100871, China}}

\begin{abstract}
We calculate the decay rate of bottomonium to two-charm quark jets
$\Upsilon \to c \bar c$ at the tree level and one-loop level
including color-singlet and color-octet $b \bar b$ annihilations. We
find that the short distance coefficient of the color-octet piece is
much larger than the color-singlet piece, and that the QCD
correction will change the endpoint behavior of the charm quark jet.
The color-singlet piece is strongly affected  by the one-loop QCD
correction. In contrast, the QCD correction to the color-octet piece
is weak. Once the experiment can measure the branching ratio and
energy distribution of the two-charm quark jets in the $\Upsilon$
decay, the result can be used to test the color octet mechanism or
give a strong constraint on the color-octet matrix elements.
\end{abstract}



\maketitle

\section{Introduction}
It is commonly believed that the heavy quark pair production and
annihilation decay can be described by perturbative Quantum
Chromodynamics (pQCD) since the heavy quarkonium mass provides a
scale that is much larger than $\Lambda_{QCD}$.  Due to its
nonrelativistic nature, the heavy quarkonium annihilation decay is
expected to be described in an effective theory, non-relativistic
Quantum Chromodynamics (NRQCD)\cite{Bodwin:1994jh}. In the NRQCD
factorization formalism, the decay of heavy quarkonium is described
by a series of annihilations of the heavy quark pair states and
corresponding long-distance matrix elements, which are scaled by the
relative velocity $v$ of quark and antiquark in the quarkonium rest
frame. The heavy quark pair states can have not only the same
quantum numbers as those of the quarkonium, but also other different
quantum numbers in color and angular momentum. In particular, the
heavy quark pair can be in a color-octet state.

The color-octet scenario seems to acquire some significant successes
in describing heavy quarkonium decay and production. But recently,
several next-to-leading order (NLO) QCD corrections for the
inclusive and exclusive heavy quarkonium production in the
color-singlet piece are found to be large and significantly relieve
the conflicts between the color-singlet model predictions and
experiments. It may imply, though inconclusively, that the
color-octet contributions in the production processes are not as big
as previously expected, and the color-octet mechanism should be
studied more carefully.

The current experimental results on inelastic $J/\psi$
photoproduction at HERA are adequately described by the NLO color
singlet piece~\cite{Kramer:1995nb}. The DELPHI data favor the NRQCD
formalism for $J/\psi$ production in $\gamma  \gamma \rightarrow
J/\psi X$, rather than the color-singlet
model\cite{Klasen:2001cu,Abdallah:2003du}. The large discrepancies
in $J/\psi$ production via double $c\bar c$ in $e^+e^-$ annihilation
at B factories between LO theoretical predictions
\cite{Braaten:2002fi,
Liu:2002wq,Hagiwara:2003cw,Yuan:1996ep,Liu:2003zr,Liu:2003jj} and
experimental results ~\cite{Abe:2002rb,Aubert:2005tj} are probably
resolved by including the higher order corrections: NLO QCD
corrections and relativistic corrections
\cite{Zhang:2005ch,Zhang:2006ay,Zhang:2008gp,
Gong:2008ce,Gong:2007db,He:2007te,bodwin06}. The NLO QCD corrections
in $J/\psi$ and $\Upsilon$ production at the Tevatron and LHC are
calculated including the color singlet
piece~\cite{Campbell:2007ws,Artoisenet:2007xi} and the color octet
piece~\cite{Gong:2008ft}. The QCD corrections to polarizations  of
$J/\psi$ and $\Upsilon$ at the Tevatron and LHC are also calculated
~\cite{Gong:2008hk,Gong:2008sn,Gong:2008ft}. The experimental data
of polarizations at the Tevatron seem to favor the NLO QCD
corrections of the color singlet piece rather than the color octet
piece. Recent developments and related topics in quarkonium physics
can be found in
Refs.~\cite{Brambilla:2004wf,Lansberg:2006dh,Lansberg:2008zm}.

In order to further test the color octet mechanism, in this paper we
calculate the rate of bottomonium decay into a charm quark pair
$\Upsilon \to c \bar c$. There have been some works on bottomonium
decays and the color octet mechanism. Fritzsch and Streng calculated
the decay rate of $\Upsilon$ into charm at leading order in
$\alpha_s$, $\Upsilon \to g gg^* \to g g c\bar
c$\cite{Fritzsch:1978ey}. Bigi and Nussinov have taken into account
the contribution of $\Upsilon \to g g^*g^* \to g c\bar c$
\cite{Bigi:1978tj}. Barbieri, Caffo, and Remiddi have calculated the
decay rates of the $P$-wave bottomonium states into charm at leading
order in $\alpha_s$ \cite{Barbieri:1979gg}. Maltoni and Petrelli
calculated the effects of color-octet contributions on the radiative
$\Upsilon$ decay \cite{Maltoni:1998nh}. Recently, Bodwin, Braaten
and Kang calculated the inclusive decay rate of $\chi_b$ into
charmed hadron in the NRQCD framework\cite{Bodwin:2007zf}. Gao,
Zhang and Chao calculated the bottomonium radiative decays to
charmonium and light mesons\cite{Gao:2006bc,Gao:2007fv}, as well as
$\Upsilon$ radiative decay to light quark jet to test the color
octet mechanism\cite{Gao:2006ak}. The S-wave quarkonium decay to
light hadrons was calculated up to order $v^4$ and $\alpha_s^3$
\cite{Petrelli:1997ge,Bodwin:2002hg}. The exclusive double
charmonium production from $\Upsilon$ decay was calculated by Jia
\cite{Jia:2007hy}. Kang, Kim, Lee and Yu have calculated the
inclusive charm production in $\Upsilon(nS)$ decay
\cite{Kang:2007uv}. And the invariant-mass distribution of $c \bar
c$ in $\Upsilon(1S) \to c\bar c + X$ was also calculated by Chung,
Kim and Lee\cite{Chung:2008yf}. And  $\eta_b$  inclusive charm decay
was calculated by Hao, Qiao and Sun \cite{Hao:2007rb}. As to
experiment, ARGUS Collaboration searched for charm production in
direct decays of the $\Upsilon(1S)$, and found $B^{dir}[\Upsilon(1S)
\to D^*(2010) ^\pm + X] < 0.019$\cite{Albrecht:1992ap}. Very
recently CLEO has searched for the $D^0$ production in direct decays
of the $\chi_{bJ}(nS)$ (n=1,2) states \cite{Briere:2008cv}. The
present investigation for the $\Upsilon$ decay to $c\bar c$ pair
will hopefully add a new contribution to the test of color-octet
mechanism in heavy quarkonium decays.

This paper is organized as follows. In Sec.~\ref{sec:TheoFram}, we
present the theoretical framework for the decay of $\Upsilon \to c
\bar c$. In Sec.~\ref{sec:singlet}, we estimate the color-singlet
contributions. In Sec.~\ref{sec:octet}, we include the color-octet
contributions. In Sec.~\ref{sec:matrixelement}, we discuss the NRQCD
matrix elements e.g. $\left\langle\Upsilon|{ \cal O}({}^3S_{1,8})|
\Upsilon \right\rangle$ and give a numerical estimation of the
color-octet contributions.
Summary and discussion are presented in Sec.~\ref{sec:summ&dis}. The
detailed and lengthy intermediate steps and formulas in the
calculation will be given in the appendices.

\section{Theoretical Framework}
\label{sec:TheoFram}%

In the framework of NRQCD, the width of $~\Upsilon$~ decay to $~c
\bar c$~ can be written as
\begin{eqnarray}
\Gamma [\Upsilon \to c \bar c]=\sum_n \hat \Gamma [b \bar b (n) \to
c \bar c] \left\langle\Upsilon|{ \cal O}(n)| \Upsilon \right\rangle,
\end{eqnarray}
where $n$ denote quantum numbers including the spin angular momentum
S, orbit angular momentum L, total angular momentum J, and the color
index $1$ or $8$. The short distance coefficients $\hat \Gamma [b
\bar b (n) \to c \bar c]$ can be calculated in pQCD, and the long
distance factors $ \left\langle\Upsilon|{ \cal O}(n)| \Upsilon
\right\rangle$ scale as definite powers of the relative velocity $v$
of quark and antiquark in the quarkonium rest
frame~\cite{Bodwin:1994jh}. For $\Upsilon$, the leading order matrix
element is $\left\langle\Upsilon|{ \cal O}({}^3S_{1,1})| \Upsilon
\right\rangle$, and there are three matrix elements that contribute
up to corrections of relative order $v^4$:~ $\left\langle\Upsilon|{
\cal O}({}^1S_{0,8})| \Upsilon \right\rangle$,
$\left\langle\Upsilon|{ \cal O}({}^3S_{1,8})| \Upsilon
\right\rangle$, and $\left\langle\Upsilon|{ \cal O}({}^3P_{J,8})|
\Upsilon \right\rangle$. The other matrix elements are of higher
order in $v$.

\begin{center}
\begin{figure}[!hbp]
\includegraphics[width=14.0cm]{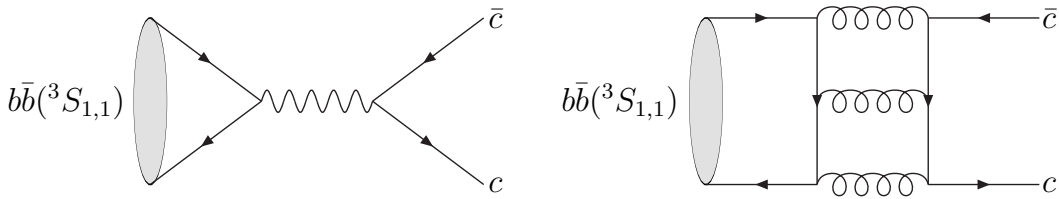}
\caption{\label{fig:bbccbarSL}Feynman diagrams for the color-singlet
decay $\Upsilon(b \bar b ({}^3S_{1,1})) \to c \bar c$ via a virtual
photon (left) and three virtual gluons (right).}
\end{figure}
\end{center}
Feynman diagrams for the color-singlet decay $\Upsilon(b \bar b
({}^3S_{1,1})) \to c \bar c$ via a virtual photon (left) and three
virtual gluons (right) are shown in Fig.~\ref{fig:bbccbarSL}. At
leading order in $\alpha_s$, the decay of the color-singlet piece
$\Upsilon(b \bar b ({}^3S_{1,1}))$ can proceed through a virtual
photon or three virtual gluons. The decay width is of order ${\cal
O}( \left(\alpha/\pi\right)^2)$  for the virtual photon, and$~{\cal
O}( \left(\alpha_s/\pi\right)^6)$ for the three-gluons. So the
single photon process is expected to be dominant, and the
contribution of three-gluon process will be roughly estimated in
Sec.~\ref{sec:singlet}. If a soft gluon is allowed to appear in the
final state, the order of $\alpha_s$ in the process can be
decreased. But the processes of $b \bar b ({}^3S_{1,1}) \to 2 g^* +g
\to c \bar c +g$ and $b \bar b ({}^3S_{1,1}) \to g^* + 2 g \to c
\bar c +2g$ are infrared (IR) finite
\cite{Fritzsch:1978ey,Bigi:1978tj}, so the phase space of the soft
gluon will bring a suppression factor:
\begin{eqnarray}
\left.\frac{d ^3 k_g}{m_b^2 k_g^0} \right|_{k_g^0 < m_b \delta _s}&
\sim &\delta _s^2,
\end{eqnarray}
where the factor of $m_b^2$ is used to balance the dimension, and
$\delta_s$ is the soft cut. The gluon is regarded as a soft gluon
when the energy of the gluon is lower than $ m_b \delta _s$. When
$\delta_s$ is set to, say, $0.2$, the corresponding energy cut is
about $1$~GeV, $~\delta _s^2$ is numerically close to
$\alpha_s/\pi$, so these soft gluon processes are relatively
suppressed and should be ignored here.

\begin{center}
\begin{figure}
\includegraphics[width=14.0cm]{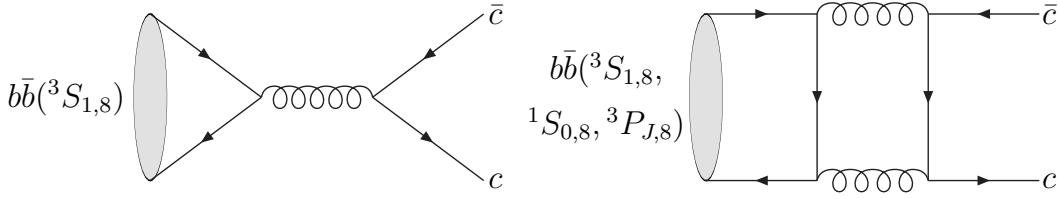}
\caption{\label{fig:bbOCTcc} Feynman diagrams for the color-octet $b
\bar b \to c \bar c$}
\end{figure}
\end{center}

The decay of the color-octet piece $b \bar b  \to c \bar c$ includes
contributions from color-octet $b \bar b$ components ${}^3S_{1,8}$,
as well as ${}^1S_{0,8}$ and $^3P_{J,8}$ in the $\Upsilon$ Fock
state expansion. Feynman diagrams for the color-octet $b \bar b \to
c \bar c$ are shown in Fig.~\ref{fig:bbOCTcc}. The leading order
decay width of $b \bar b({}^3S_{1,8}) \to c \bar c$ is of order
${\cal O}(\alpha_s^2/\pi^2)$, while processes $b \bar b
({}^1S_{0,8},{}^3P_{J,8}) \to c \bar c$ can only proceed via a loop,
and the corresponding decay widths are of order ${\cal
O}(\alpha_s^4/\pi^4)$. Moreover, since $\left\langle\Upsilon|{ \cal
O}({}^3S_{1,8})| \Upsilon \right\rangle \sim \left\langle\Upsilon|{
\cal O}({}^1S_{0,8})| \Upsilon \right\rangle \sim \frac{
\left\langle\Upsilon|{ \cal O}({}^3P_{J,8})| \Upsilon
\right\rangle}{m_b^2} \sim v^4 \left\langle\Upsilon|{ \cal
O}({}^3S_{1,1})| \Upsilon \right\rangle$ according to the velocity
scaling rule, the contributions of $\Upsilon(b \bar b
({}^1S_{0,8},{}^3P_{J,8})) \to c \bar c$ can be neglected.

The color-singlet and color-octet contributions will be discussed
respectively in the next two sections.

\section{Color-singlet piece $b \bar b ({}^3S_{1,1}) \to c
\bar c$} \label{sec:singlet}

The amplitude of color-singlet piece $b \bar b ({}^3S_{1,1}) \to c
\bar c$ can be written as~\cite{Gao:2006ak,Gao:2007fv}:
\begin{eqnarray}
\label{amp2}   {\cal A}\Big(b\bar{b}({}^{3}S_{1, 1}(2p_b)\rightarrow
c(p_c)+ \bar{c}(p_{\bar c})\Big)&=& \sqrt{\frac{
\left\langle\Upsilon|{ \cal O}({}^3S_{1,1})| \Upsilon
   \right\rangle
}{2N_c}} \sum\limits_{L_{\Upsilon z}
S_{\Upsilon z} }\sum\limits_{s_1s_2 }\sum\limits_{jk}
\nonumber\\
&&\hspace{-3.0cm}\times \langle 1 \mid \bar{3}k;3j \rangle \langle J_\Upsilon
J_{\Upsilon z} \mid L_\Upsilon L_{\Upsilon z };S_\Upsilon
S_{\Upsilon z} \rangle \langle S_\Upsilon S_{\Upsilon z} \mid
s_1;s_2 \rangle \nonumber\\
&&\hspace{-3.0cm}\times{\cal A}\Big(b_j(p_b)+\bar{b}_k(p_b)\rightarrow
c_l(p_c)+\bar{c}_i(p_{\bar c})\Big),
\end{eqnarray}
where $\langle 1 \mid \bar{3}k;3j \rangle=\delta_{jk}/\sqrt{N_c}$, $
\langle S_\Upsilon S_{\Upsilon z} \mid s_1;s_2 \rangle$, and $
\langle J_\Upsilon J_{\Upsilon z} \mid L_\Upsilon L_{\Upsilon z
};S_\Upsilon S_{\Upsilon z} \rangle$ are respectively the
color-SU(3), spin-SU(2), and angular momentum Clebsch-Gordan
coefficients for $b\bar{b}$ pairs projecting on appropriate bound
states $\Upsilon$. ${\cal A}(b_j(p_b)+\bar{b}_k(p_b)\rightarrow
c_l(p_c)+\bar{c}_i(p_{\bar c}))$ is the amplitude of the process
$b_j(p_b)+\bar{b}_k(p_b) \rightarrow c_l(p_c)+\bar{c}_i(p_{\bar
c})$. In the calculation, we use {\tt
FeynArts}~\cite{Kublbeck:1990xc,Hahn:2000kx} to generate Feynman
diagrams and amplitudes, {\tt FeynCalc}~\cite{Mertig:1990an} for the
tensor reduction, and {\tt LoopTools}~\cite{Hahn:1998yk} for the
numerical evaluation of the IR-safe integrals.

The spin projection operators $P_{SS_z}(p,q)$ which describe
quarkonium production  are expressed in terms of quark and
anti-quark spinors as\cite{Kuhn:1979bb,Guberina:1980dc}:
\begin{eqnarray} P_{SS_z}(p,q)\!=\!\!\!\sum_{s_1,s_2}\!\!u(\frac{p}{2}\!+\!q,\!s_2)
 \bar v(\!\frac{p}{2}\!-\!q,\!s_1) \!\langle s_1;\!s_2|SS_z\!\rangle.
\end{eqnarray}
For the ${}^3S_1$ state, it is
\begin{eqnarray}
P_{1S_Z}(2p_b,0)&=&\frac{1}{2\sqrt{2m_b}}(2\not{\!
p}_b+2m)\not{\epsilon}(S_z).
\end{eqnarray}
And the spin projection operators which describe the annihilation of
quarkonium are the complex conjugate of the corresponding operators
for production.

The leading order (LO) color-singlet decay $b \bar b ({}^3S_{1,1})
\to \gamma^*\to c \bar c$ is shown in Fig.~\ref{fig:bbccbarSL}, and
the corresponding  LO  decay width is
\begin{eqnarray}\label{eq:LOSOcc}
\Gamma_{LO}[\Upsilon({}^3S_{1,1}) \to c \bar c]=\frac{4\pi \alpha^2
\sqrt{1-r^2} \left(2
   +r^2\right)   }{81 m_b^2}
   \left\langle\Upsilon|{ \cal O}({}^3S_{1,1})| \Upsilon
   \right\rangle,
\end{eqnarray}
where $r^2=m_c^2/m_b^2$. This result is consistent with
Ref.\cite{Chung:2008yf,Kang:2007uv}. Comparing it with the leptonic
width
\begin{eqnarray}
\Gamma_{LO}[\Upsilon \to e^+e^-]=\frac{2
 \pi \alpha^2  }{27 m_b^2}
   \left\langle\Upsilon|{ \cal O}({}^3S_{1,1})| \Upsilon
   \right\rangle,
\end{eqnarray}
  we can get
\begin{eqnarray}\label{eq:LOSOcccc}
\Gamma_{LO}[\Upsilon({}^3S_{1,1}) \to c \bar
c]=~\frac{4}{3}~\Gamma_{LO}[\Upsilon \to e^+e^-]\times \left(1+
{\cal O}\left(r^2\right)\right),
\end{eqnarray}
where the factor $4/3$ comes from the charm quark charge and color
factor, and $~r^2\sim 10^{-2}$. If we set $~m_b=4.7$~GeV and
$~m_c=1.5$~GeV, the LO decay branching ratio is
\begin{eqnarray}
B_{LO}[\Upsilon({}^3S_{1,1}) \to c \bar c]&=&1.33B_{LO}[\Upsilon \to
e^+e^-]=3.2\%,
\end{eqnarray}
where $B[\Upsilon \to e^+e^-]=(2.38\pm0.11)\%$ is used according to
the PDG 2006 Version\cite{Yao:2006px}.

\begin{figure}
\begin{center}
\includegraphics[width=14.0cm]{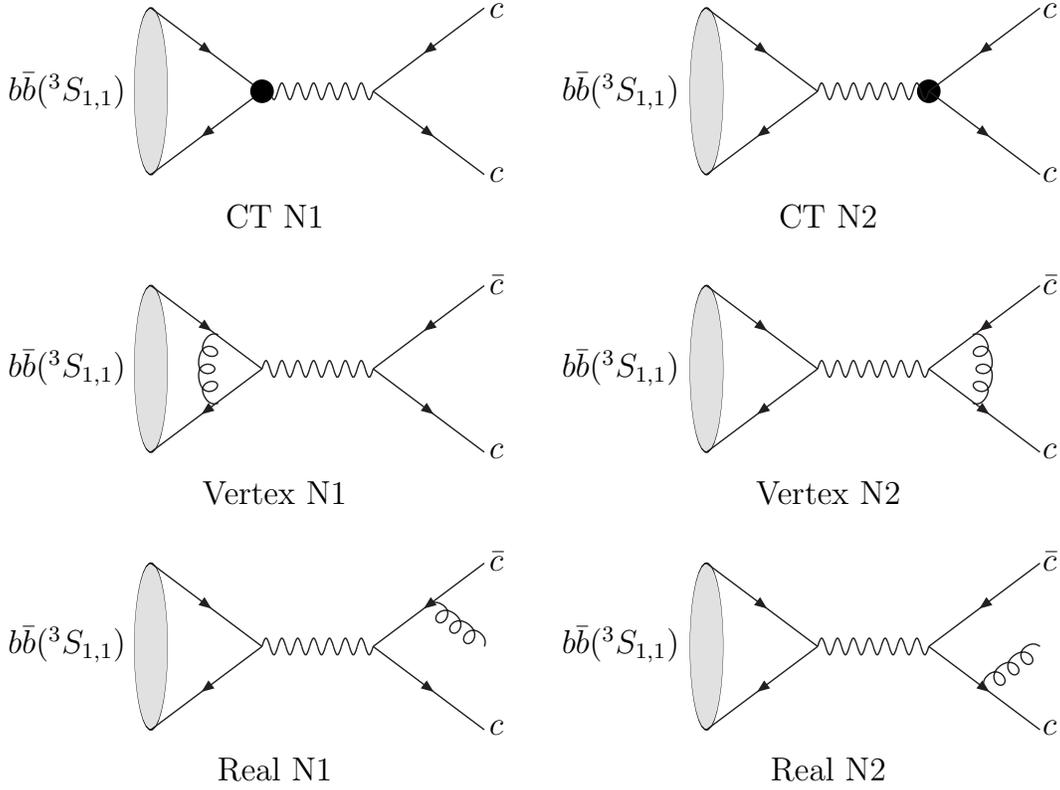}
\caption{\label{bbccSLLoop} Feynman diagrams for one-loop QCD
corrections with counter terms for $~b \bar b ({}^3S_{1,1}) \to
\gamma^*\to c \bar c$}
\end{center}
\end{figure}

We next consider the QCD radiative corrections. The Feynman diagrams
of one-loop virtual corrections and counter terms are shown in
Fig.~\ref{bbccSLLoop}. The renormalization of heavy quark wave
function should appear. The on-mass-shell (OS) scheme is chosen for
$Z_{2b}$ and $Z_{2c}$\cite{Zhang:2005ch}:
\begin{eqnarray}\label{renDefrenccbb}
\delta Z_{2b}^{\rm OS}&=&-C_F\frac{\alpha_s}{4\pi}
\left[\frac{1}{\epsilon_{\rm UV}}+\frac{2}{\epsilon_{\rm IR}}
-3\gamma_E+3\ln\frac{4\pi\mu^2}{m_b^2}+4\right]+\mathcal
{O}(\alpha_s^2),
\nonumber\\
\delta Z_{2c}^{\rm OS}&=&-C_F\frac{\alpha_s}{4\pi}
\left[\frac{1}{\epsilon_{\rm UV}}+\frac{2}{\epsilon_{\rm IR}}
-3\gamma_E+3\ln\frac{4\pi\mu^2}{m_c^2}+4\right]+\mathcal
{O}(\alpha_s^2),
\end{eqnarray}
where $\mu$ is the renormalization scale, $\gamma_E$ is the Euler's
constant. In this scheme, we need not calculate the correction of
external quark legs. We employ the two-loop formula for
$\alpha_s(\mu)$,
\begin{equation}
\frac{\alpha_s(\mu)}{4\pi}=\frac{1}{\beta_0L} -\frac{\beta_1\ln
L}{\beta_0^3L^2}, \label{eq:as}
\end{equation}
where $L=\ln\left(\mu^2/\Lambda_{\rm QCD}^2\right)$, and
$\beta_1=(34/3){C_A}^2-4 C_F T_F n_f-(20/3)C_A T_F n_f$ is the
two-loop coefficient of the QCD beta function.

The correction to $\Upsilon(b \bar b ({}^3S_{1,1})) \to \gamma^*$
gives a factor of $-\frac{16 \alpha_s}{3\pi}$ at ${\cal O}
(\alpha_s)$. The other part is the correction to $ \gamma^* \to c
\bar c$. If we set $m_c=0$, then the combined total correction
becomes the correction to $R$, the ratio of cross section of
$e^+e^-\to light~ hadrons$ to that of $e^+e^-\to \mu^+\mu^-$. It
give a factor of $\frac{ \alpha_s}{\pi}$ at ${\cal O} (\alpha_s)$.

Compared with it, the  leptonic width of $\Upsilon$ at ${\cal O}
(\alpha_s)$ becomes the known result:
\begin{eqnarray}\label{eq:NLOUpsEE}
\Gamma_{NLO}[\Upsilon \to e^+e^-]=\frac{2
 \pi \alpha^2  }{27 m_b^2}\left(1-\frac{16 \alpha_s}{3 \pi}\right)
   \left\langle\Upsilon|{ \cal O}({}^3S_{1,1})| \Upsilon
   \right\rangle.
\end{eqnarray}
If the parameters are chosen as  $m_b=4.7$~GeV, $m_c=1.5$~GeV,
and $\alpha_s=0.220$, then the branching ratio is
\begin{eqnarray}
B_{NLO}[\Upsilon({}^3S_{1,1}) \to c \bar c+X]&=&1.6 B_{NLO}[\Upsilon
\to e^+e^-]\approx 3.8\%.
\end{eqnarray}
When the emitted gluon energy is large, it would form a jet. So  a
cut to the gluon energy  should be introduced to distinguish between
$c\bar c$ and $c\bar cg$ final states. If $E_g< m_b \times
\delta_s$, the gluon is considered as soft and the final state is
$c\bar c$. Otherwise, when $~E_g> m_b \times \delta_s$, the final
state is $c\bar cg$. If we set $\delta_s =0.15$, then the branching
ratio is
\begin{eqnarray}
B_{NLO}[\Upsilon({}^3S_{1,1}) \to c \bar c]&\approx &1.4\%.
\end{eqnarray}
If we set $\delta_s =0.10$ and $0.20$, then the branching ratio is
about $5.4\times 10^{-3}$ and $2.0\%$ respectively.

For the color-singlet piece, the contribution of
$\Upsilon({}^3S_{1,1}) \to 3 g^* \to c \bar c$, which is shown on
the right hand side in Fig.~\ref{fig:bbccbarSL}, has not been
calculated so far, and we may have a rough estimate for it.
We can use $\Upsilon({}^3S_{1,1}) \to 3 g$ to give an order of
magnitude estimate for the contribution of $\Upsilon({}^3S_{1,1})
\to 3 g^* \to c \bar c$. We have the following order of magnitude
estimates:
\begin{eqnarray}\label{eq:orderestimate}
B[\Upsilon({}^3S_{1,1}) \to l^+l^-] &\propto&
\left(\frac{\alpha}{\pi}\right)^2 ,\nonumber \\
B[\Upsilon({}^3S_{1,1}) \to 3 g ] &\propto&
\left(\frac{\alpha_s}{\pi}\right)^{3} ,\nonumber \\
B[\Upsilon({}^3S_{1,1}) \to g g g^* \to ggc\bar c  ] &\propto&
\left(\frac{\alpha_s}{\pi}\right)^{4} ,\nonumber \\
B[\Upsilon({}^3S_{1,1}) \to 3 g^* \to c \bar c ]&\propto&
\left(\frac{\alpha_s}{\pi}\right)^6.
\end{eqnarray}
Comparing the leptonic width with three-gluon width, we get
\begin{eqnarray}
\frac{\Gamma[\Upsilon({}^3S_{1,1}) \to
l^+l^-]}{\Gamma[\Upsilon({}^3S_{1,1}) \to 3 g ] } &\sim &
\frac{\left(\frac{\alpha}{\pi}\right) ^2 }
{\left(\frac{\alpha_s}{\pi}\right)^3}\sim 0.016,
\end{eqnarray}
which is about one half of  the  experimental value of
$\frac{\Gamma[\Upsilon({}^3S_{1,1}) \to
l^+l^-]}{\Gamma[\Upsilon({}^3S_{1,1}) \to 3 g ] }\approx
0.03$\cite{Yao:2006px}, and the closeness of this estimate to the
data may suggest that the naive estimate could make sense. Comparing
$\Upsilon({}^3S_{1,1}) \to 3 g^* \to c \bar c$ with
$\Upsilon({}^3S_{1,1}) \to 3 g$ and $\Upsilon({}^3S_{1,1}) \to
\gamma^* \to c \bar c$, we can get
\begin{eqnarray}\label{eq:ccratio}
\frac{\Gamma[\Upsilon({}^3S_{1,1}) \to 3 g^*
 \to c \bar c]}{\Gamma[\Upsilon({}^3S_{1,1}) \to 3 g ] } &\sim &
\left(\frac{\alpha_s}{\pi}\right)^3 \approx 3\times 10^{-4},
\end{eqnarray}

\begin{eqnarray}\label{eq:cc3gandgamma}
\frac {\Gamma[\Upsilon({}^3S_{1,1}) \to 3 g^* \to c \bar c]
}{\Gamma[\Upsilon({}^3S_{1,1}) \to \gamma^* \to c \bar c]}&\sim &
\frac{\alpha_s^6}{\alpha^2 \pi^4} \approx 0.02.
\end{eqnarray}
From the estimates given in  Eq.(\ref{eq:ccratio}) and
Eq.(\ref{eq:cc3gandgamma}), we see that the contribution of
$\Upsilon({}^3S_{1,1}) \to 3 g^* \to c \bar c$ is very small and
much smaller than that of $\Upsilon({}^3S_{1,1}) \to \gamma^* \to c
\bar c$. Even if the contribution of $\Upsilon({}^3S_{1,1}) \to 3
g^* \to c \bar c$ is underestimated by an order of magnitude in
Eq.(\ref{eq:ccratio}) and Eq.(\ref{eq:cc3gandgamma}), we could still
expect that for the decay $\Upsilon({}^3S_{1,1}) \to c \bar c$ the
QED contribution is dominant. Another useful example is the decay
rate $\Gamma[\Upsilon({}^3S_{1,1}) \to g g c \bar c ]$, which is of
higher order in $\alpha_s$ than $\Gamma[\Upsilon({}^3S_{1,1}) \to
ggg]$, and is given in Ref.\cite{Kang:2007uv}. From their estimate
we can get $\frac{\Gamma[\Upsilon({}^3S_{1,1}) \to g g c \bar c ]
}{\Gamma[\Upsilon({}^3S_{1,1}) \to ggg]}=0.029$, which is also of
the same order of magnitude as, but even smaller than, our naive
estimate.
\begin{eqnarray}
 \frac{\Gamma[\Upsilon({}^3S_{1,1}) \to g g c \bar c ] }
 {\Gamma[\Upsilon({}^3S_{1,1}) \to g g g]} &\sim
&\left(\frac{\alpha_s}{\pi}\right) \approx 0.071.
\end{eqnarray}
This might imply that the naive estimates given in
Eq.(\ref{eq:orderestimate}) as well as in Eq.(\ref{eq:cc3gandgamma})
might be tenable in estimating the rates of higher order processes
by order of magnitude. So, based on the rough estimate given in
Eq.(\ref{eq:cc3gandgamma}) for the contributions of the
color-singlet piece to the $\Upsilon({}^3S_{1,1}) \to c \bar c$
process, we assume that, as an approximation, the contribution of
$\Upsilon({}^3S_{1,1}) \to 3 g^* \to c \bar c$ can be neglected, and
only $\Upsilon({}^3S_{1,1}) \to\gamma^* \to c \bar c$ will be taken
into consideration.



\section{Color octet piece $b \bar b ({}^3S_{1,8}) \to c
\bar c$} \label{sec:octet}

The amplitude of color-octet piece $b \bar b ({}^3S_{1,8}) \to c
\bar c$ can be written as~\cite{Gao:2006ak,Gao:2007fv}
\begin{eqnarray}
\label{amp2}   {\cal A}\Big(b\bar{b}({}^{3}S_{1, 8}(2p_b)\rightarrow
c(p_c)+ \bar{c}(p_{\bar c})\Big)&=& \sqrt{ \left\langle\Upsilon|{
\cal O}({}^3S_{1,8})| \Upsilon
   \right\rangle
} \sum\limits_{L_{\Upsilon z} S_{\Upsilon z} }\sum\limits_{s_1s_2
}\sum\limits_{jk}
\nonumber\\
&&\hspace{-3.0cm}\times \langle 8~a \mid \bar{3}k;3j \rangle \langle
J J_{ z} \mid L L_{ z };S S_{ z} \rangle \langle S S_{ z} \mid
s_1;s_2 \rangle \nonumber\\
&&\hspace{-3.0cm}\times{\cal
A}\Big(b_j(p_b)+\bar{b}_k(p_b)\rightarrow c_l(p_c)+\bar{c}_i(p_{\bar
c})\Big),
\end{eqnarray}
where $\langle 8~a \mid \bar{3}k;3j \rangle=\sqrt 2 T^a_{jk}$, and
other expressions are similar to the color-singlet piece.

The Born diagram of $b \bar b ({}^3S_{1,8}) \to c \bar c$ is shown
in Fig.~\ref{fig:bbOCTcc}. It is also calculated in
Ref.~\cite{Bodwin:2007zf}. The leading order width is
\begin{eqnarray}\label{eq:LOOCTcc}
\Gamma_{LO}[\Upsilon({}^3S_{1,8}) \to c \bar c]= \frac{\alpha_s^2
\sqrt{1-r^2} \left(2
  +r^2\right)  \pi }{6 m_b^2}
   \left\langle\Upsilon|{ \cal O}({}^3S_{1,8})| \Upsilon
   \right\rangle.
\end{eqnarray}

\begin{figure}
\begin{center}
\includegraphics[width=14.0cm]{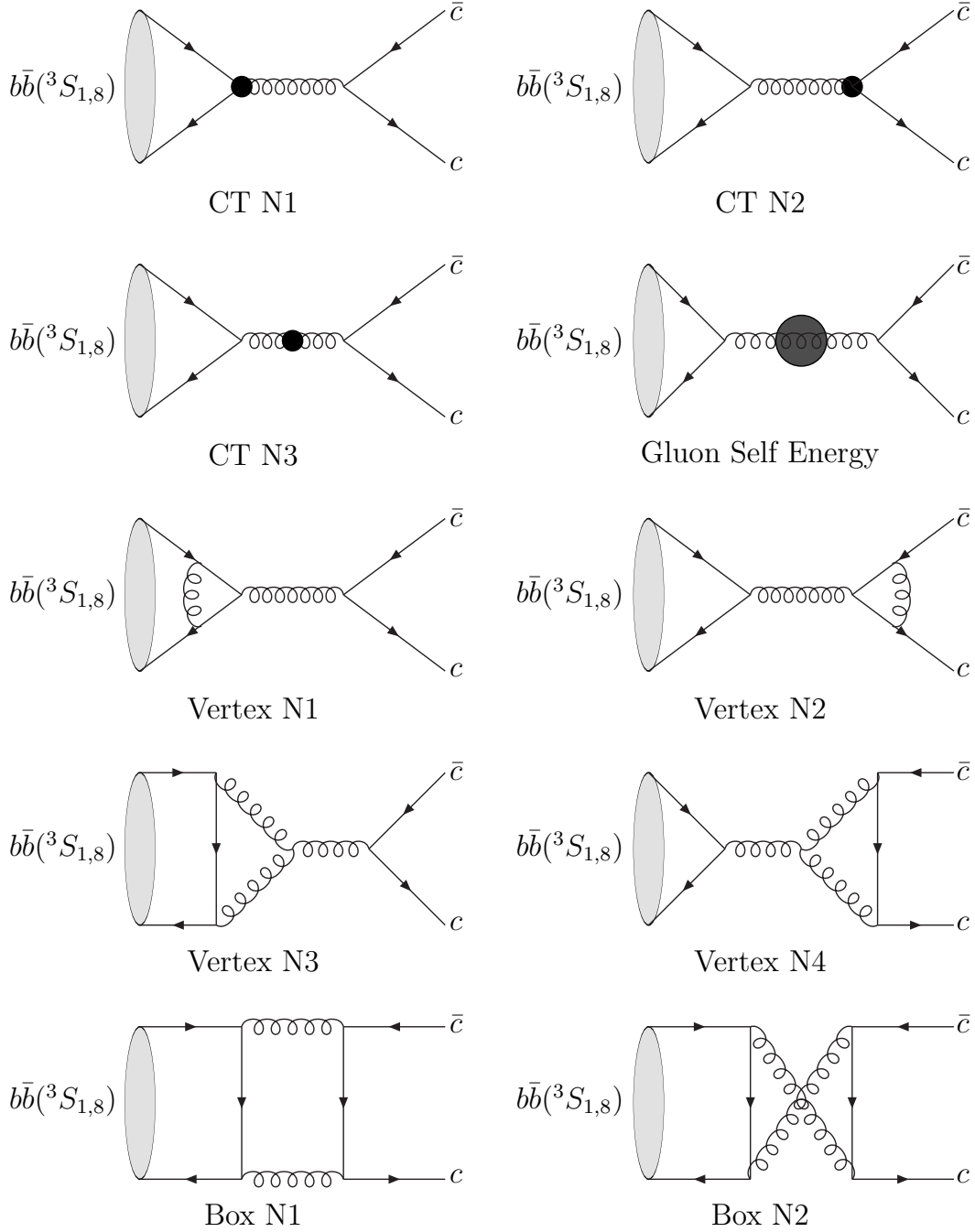}
\caption{\label{bbccOCTLOOP} Feynman diagrams for next-to-leading
order virtual corrections with counter terms in the color-octet
piece $b \bar b ({}^3S_{1,8}) \to c \bar c$}
\end{center}
\end{figure}

\begin{figure}
\begin{center}
\includegraphics[width=14.0cm]{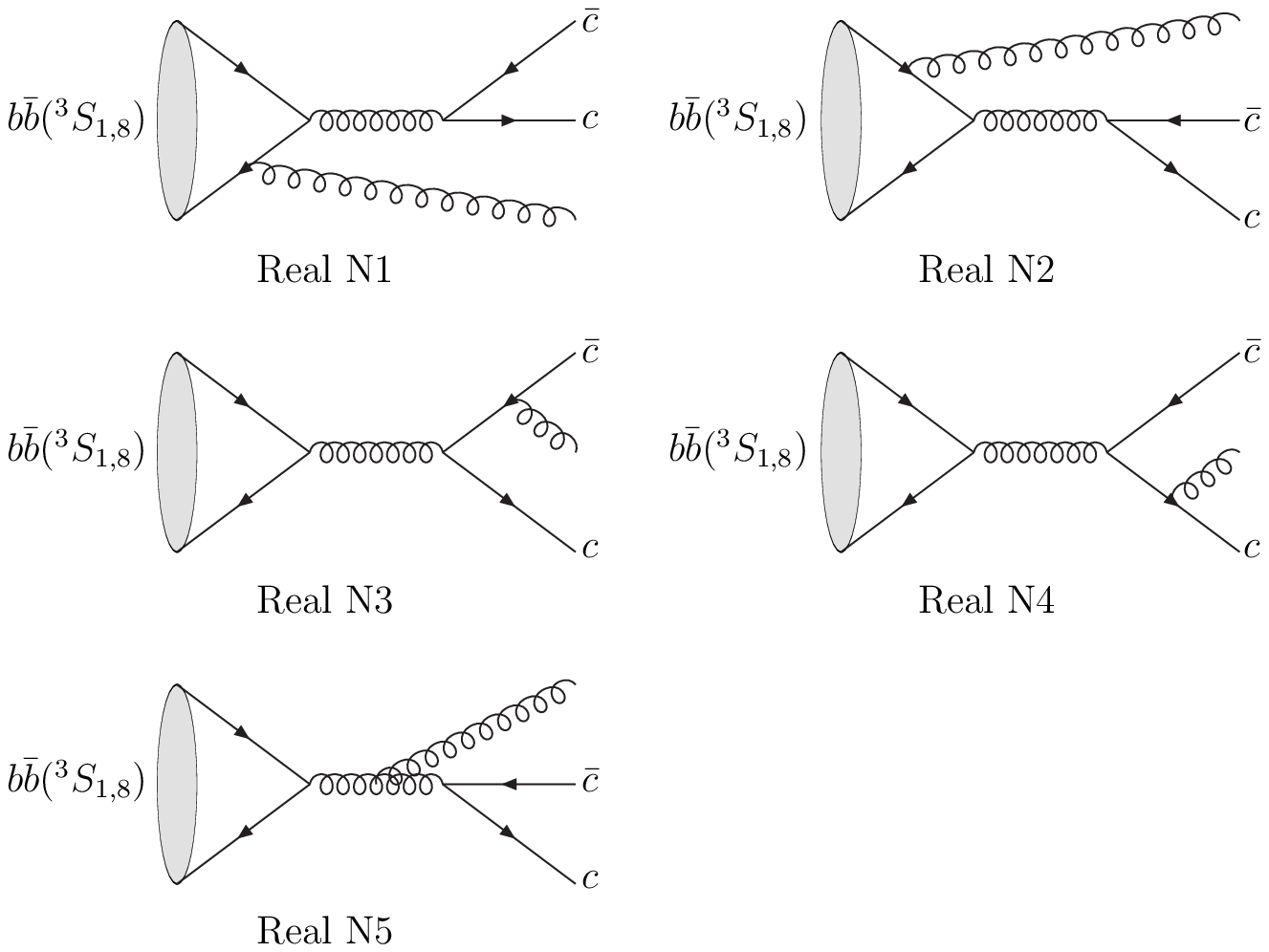}
\caption{\label{bbccOCTREAL}Feynman diagrams for next-to-leading
order real corrections in the color-octet piece $b \bar b
({}^3S_{1,8}) \to c \bar c$}
\end{center}
\end{figure}

We further calculate the next-to-leading order (NLO) corrections.
The Feynman diagrams for NLO virtual corrections with counter terms
in the color-octet piece $b \bar b ({}^3S_{1,8}) \to c \bar c$ are
shown in Fig.~\ref{bbccOCTLOOP}. The Feynman diagrams for  NLO real
corrections in the color-octet piece $b \bar b ({}^3S_{1,8}) \to c
\bar c$  are shown in Fig.~\ref{bbccOCTREAL}. The renormalization of
heavy quark wave function, gluon wave function, and coupling
constant should appear here. $Z_{2b}$ and $Z_{2c}$ are given in
Eq.~(\ref{renDefrenccbb}). For $Z_{3}$ and $Z_{g}$, we choose the
modified minimal-subtraction ($~\overline{\rm MS}$~)
scheme\cite{Zhang:2005ch}:
\begin{eqnarray}\label{renDefren}
\delta Z_3^{\overline{\rm MS}}&=&\frac{\alpha_s}{4\pi}
(\beta_0-2C_A) \left[\frac{1}{\epsilon_{\rm UV}} -\gamma_E +
\ln(4\pi)\right]+\mathcal
{O}(\alpha_s^2), \nonumber\\
  \delta Z_g^{\overline{\rm MS}}&=&-\frac{\beta_0}{2}\,
  \frac{\alpha_s}{4\pi}
  \left[\frac{1}{\epsilon_{\rm UV}} -\gamma_E + \ln(4\pi)
  \right]+\mathcal
{O}(\alpha_s^2).
\end{eqnarray}

The parameters are chosen as $m_b=4.7$~GeV, $m_c=1.5$~GeV, $n_f=4$,
$\Lambda_{QCD}^{(4)}=338$~MeV, $\mu=m_b$, and then $\alpha_s=0.220$.
So we can get the leading order result:
\begin{eqnarray}
B_{LO}[\Upsilon({}^3S_{1,8}) \to c \bar c]&=&42 \times \frac{
\left\langle\Upsilon|{ \cal O}({}^3S_{1,8})| \Upsilon
   \right\rangle}{\rm GeV^3}.
\end{eqnarray}
The total NLO result is
\begin{eqnarray}\label{eq:bbccOCTNLO}
B_{NLO}[\Upsilon({}^3S_{1,8}) \to c \bar c+X]&=&53 \times \frac{
\left\langle\Upsilon|{ \cal O}({}^3S_{1,8})| \Upsilon
   \right\rangle}{\rm GeV^3}.
\end{eqnarray}
If we set the soft cut $~\delta_s =0.15$, then the NLO  result is
\begin{eqnarray}
B_{NLO}[\Upsilon({}^3S_{1,8}) \to c \bar c]&=&41 \times \frac{
\left\langle\Upsilon|{ \cal O}({}^3S_{1,8})| \Upsilon
   \right\rangle}{\rm GeV^3}.
\end{eqnarray}
If we set the soft cut  $\delta_s =0.10$ and $0.20$, then the
branching ratio is $37 \times \frac{ \left\langle\Upsilon|{ \cal
O}({}^3S_{1,8})| \Upsilon
   \right\rangle}{\rm GeV^3}$ and $44 \times \frac{
\left\langle\Upsilon|{ \cal O}({}^3S_{1,8})| \Upsilon
   \right\rangle}{\rm   GeV^3}$ respectively.

From the above expressions, we see that the short-distance
coefficient for this color-octet process is large, and this
color-octet process may make a significant contribution to the
$\Upsilon$ decay to two-charm quark jet. The numerical estimate will
be given in the next section.

The color-octet pieces $b \bar b ({}^3P_{J,8}) $ and  $b \bar b
({}^1S_{0,8})$ also contribute to the charm quark jet production
through $c \bar c g$, where the gluon is soft. The $b \bar b
({}^3P_{J,8})\to c \bar c g $ is IR divergent, and it should be
absorbed into the matrix element $ \left\langle\Upsilon|{ \cal
O}({}^3S_{1,8})| \Upsilon
   \right\rangle$  \cite{Bodwin:1994jh}:
\begin{eqnarray}
\left\langle\Upsilon\left|{\cal
O}\left({}^3\!S_{1,8}\right)\right|\Upsilon\right\rangle_1
&=&\left\langle\Upsilon\left|{\cal
O}^H\left({}^3\!S_{1,8}\right)\right|\Upsilon\right\rangle_0
\left[1+\left(C_F-\frac{C_A}{2}\right)\frac{\pi\alpha_s}{2v}\right]
+\frac{4\alpha_s}{3\pi
m_b^2}\left(\frac{4\pi\mu^2}{\lambda^2}\right)^\epsilon
\nonumber\\
&&{}\times\exp(-\epsilon\gamma_E)\left(\frac{1}{\epsilon_{\rm
UV}}-\frac{1}{\epsilon_{\rm IR}}\right) \sum_{J=0}^2
B_F\left\langle\Upsilon\left|{\cal
O}\left({}^3\!P_{J,8}\right)\right|\Upsilon\right\rangle,
\label{eq:reg}
\end{eqnarray}
where the Coulomb term of $\left\langle\Upsilon\left|{\cal
O}\left({}^3\!S_{1,8}\right)\right|\Upsilon\right\rangle_1$ is
canceled by the virtual correction, the IR divergent  terms is
canceled by $b \bar b ({}^3P_{J,8})\to c \bar c g $, and the UV
divergent term gives the running of matrix element. If we choose the
matrix element renormalization scale as $m_b$, then we find the
branching ratio of  $b \bar b ({}^3P_{J,8}) $ and  $b \bar b
({}^1S_{0,8})$ decays into $c \bar c g$ at order of $\alpha_s^3$ to
be
\begin{eqnarray}\label{eq:bbccOCTcPlus}
B[\Upsilon({}^1S_{0,8}) \to c \bar c+X]&=&2.8\times \frac{
\left\langle\Upsilon|{ \cal O}({}^1S_{0,8})| \Upsilon
   \right\rangle}{\rm GeV^3}, \nonumber \\
   B[\Upsilon({}^3P_{J,8}) \to c \bar c+X]&=&0.61\times \frac{
\left\langle\Upsilon|{ \cal O}({}^3P_{0,8})| \Upsilon
   \right\rangle}{\rm GeV^5}.
\end{eqnarray}
Since $\left\langle\Upsilon|{ \cal O}({}^1S_{0,8})| \Upsilon
\right\rangle$, $\left\langle\Upsilon|{ \cal O}({}^3S_{1,8})|
\Upsilon \right\rangle$, and $\left\langle\Upsilon|{ \cal
O}({}^3P_{J,8})| \Upsilon \right\rangle/m_b^2$ are of the same order
, we can ignore the contribution of  $b \bar b ({}^3P_{J,8}) $ and
$b \bar b ({}^1S_{0,8})$, as compared with the $b \bar b
({}^3S_{1,8})$ contribution given in Eq.(\ref{eq:bbccOCTNLO}).

\begin{figure}
\begin{center}
\includegraphics[width=14.0cm]{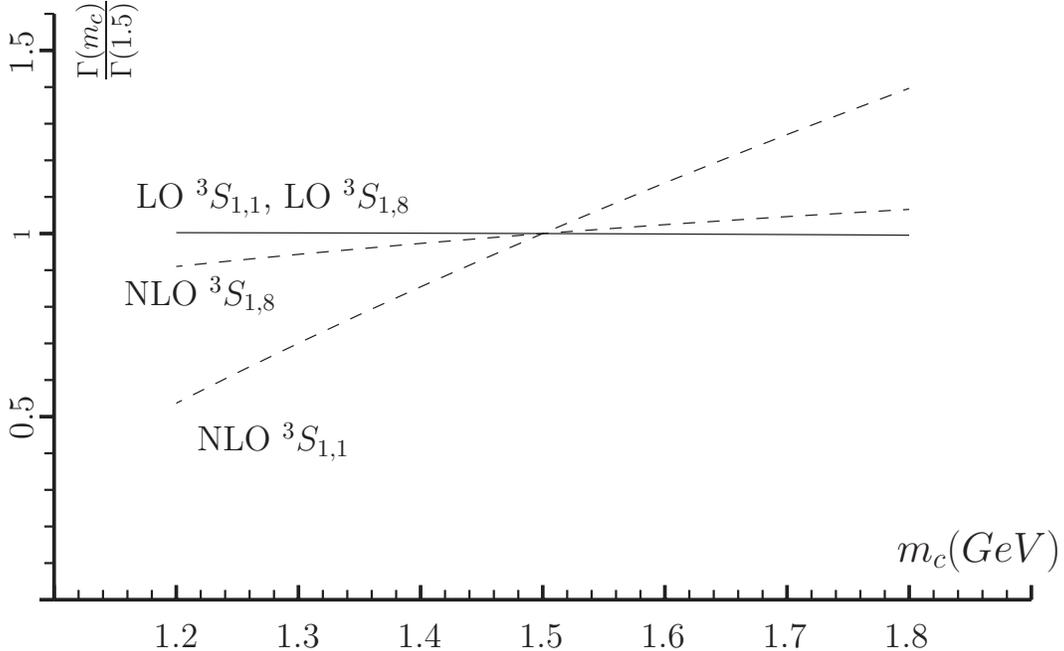}
\caption{\label{fig:depMC}
Decay widths of the color-singlet piece and color-octet piece $ b
\bar b ({}^3S_1) \to c \bar c$ rescaled by the corresponding value
at $m_c=1.5~$GeV as functions of the charm quark mass $m_c$. Here
$\Lambda=0.338 {\rm GeV}$, $ m_b=4.7~{\rm GeV} $, $\mu=m_b$, and the
soft cut $\delta_s=0.15$. LO means leading order, and NLO means
next-to-leading order. ${}^3S_{1,1}$ means the ratio $\Gamma[ b \bar
b ({}^3S_{1,1} \to c \bar c](m_c)/\Gamma[ b \bar b ({}^3S_{1,1} \to
c \bar c](m_c=1.5~{\rm GeV})$ in the color-singlet piece, and
${}^3S_{1,8}$ means the corresponding ratio in the color-octet
piece. }
\end{center}
\end{figure}

The dependence of the leading order and next-to-leading order decay
widths in the color-singlet and color-octet pieces $b\bar b \to c
\bar c$ on the charm quark is shown in Fig.~\ref{fig:depMC}. The
dependence of LO result on the charm quark mass is weak and the same
for the color-singlet and color-octet pieces. The reason can be
found in Eq.~(\ref{eq:LOSOcc}) and Eq.~(\ref{eq:LOOCTcc}). If we
choose $m_c=1.5 \pm 0.2~$GeV, the ratio is about $1 \pm 0.003$ at LO
in $\alpha_s$, $1^{+0.27}_{-0.30}$ at NLO for the color-singlet
piece, and $1^{+0.046}_{-0.057}$ at NLO for the color-octet piece.

\begin{figure}
\begin{center}
\includegraphics[width=14.0cm]{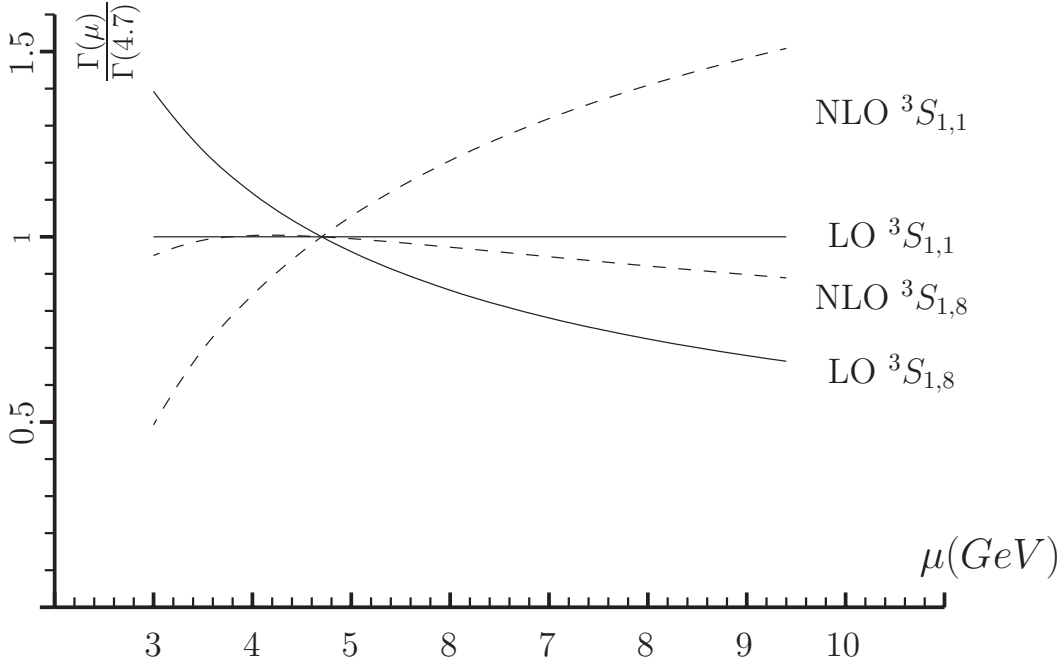}
\caption{\label{fig:depmu}
Decay widths of the color-singlet piece and color-octet piece $ b
\bar b ({}^3S_1) \to c \bar c$ rescaled by the corresponding value
at $\mu=m_b$ as functions of the renormalization scale $\mu$. Here
$\Lambda=0.338 {\rm GeV}$, $ m_b=4.7~{\rm GeV} $, $m_c=1.5~$GeV, and
the soft cut $\delta_s=0.15$. LO means leading order, and NLO means
next-to-leading order. ${}^3S_{1,1}$ means the ratio of $\Gamma[ b
\bar b ({}^3S_{1,1} \to c \bar c](\mu)/\Gamma[ b \bar b ({}^3S_{1,1}
\to c \bar c](\mu=m_b)$ in the color-singlet piece, and
${}^3S_{1,8}$ means the corresponding ratio in the color-octet
piece. }
\end{center}
\end{figure}

The dependence of the leading order and next-to-leading order decay
widths in the color-singlet and color-octet pieces $b\bar b \to c
\bar c$ on the renormalization scale $\mu$ is shown in
Fig.~\ref{fig:depmu}. The LO color-singlet result is independent of
the renormalization scale. As it is shown in the curve of NLO
${}^3S_{1,8}$, we choose $\mu=m_b$ for the “principle of minimum
sensitivity”(PMS) \cite{Stevenson:1980du}.

\begin{figure}
\begin{center}
\includegraphics[width=14.0cm]{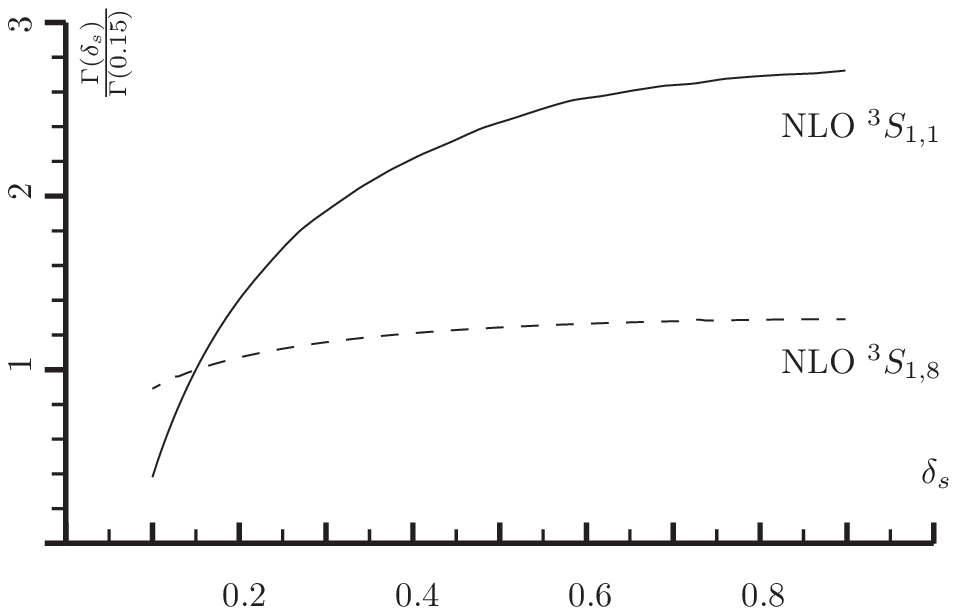}
\caption{\label{fig:depCUT} Decay widths of the color-singlet piece
and color-octet piece $ b \bar b ({}^3S_1) \to c \bar c$ rescaled by
the corresponding value at $\delta_s=0.15$ as functions of the soft
cut $\delta_s$. Here $\Lambda=0.338 {\rm GeV}$, $ m_b=4.7~{\rm GeV}
$, $ \mu=m_b$, and $m_c=1.5~$GeV. NLO means next-to-leading order.
${}^3S_{1,1}$ means the ratio of $\Gamma[ b \bar b ({}^3S_{1,1} \to
c \bar c](\delta_s)/\Gamma[ b \bar b ({}^3S_{1,1} \to c \bar
c](\delta_s=0.15)$, and ${}^3S_{1,8}$ means the corresponding ratio
in the color-octet piece. }
\end{center}
\end{figure}

The dependence of the next-to-leading order decay widths in the
color-singlet and color-octet pieces $b\bar b \to c \bar c$ on the
soft cut $\delta_s$ is shown in Fig.~\ref{fig:depCUT}. The LO result
is independent of the soft cut $\delta_s$. The NLO color-singlet
result is rather sensitive to $\delta_s$, whereas the NLO
color-octet result is insensitive to $\delta_s$.

\section{Color octet matrix elements}\label{sec:matrixelement}
The color-singlet matrix element $\left\langle\Upsilon|{ \cal
O}({}^3S_{1,1})| \Upsilon\right\rangle$ can be extracted from the
$\Upsilon$ leptonic decay width. Using Eq.~(\ref{eq:NLOUpsEE}), we
get
\begin{eqnarray}\left\langle\Upsilon|{ \cal O}({}^3S_{1,1})| \Upsilon
\right\rangle=3.8~ {\rm GeV^3}.
\end{eqnarray}

On the other hand, large uncertainty is related to the color octet
matrix element $\left\langle\Upsilon|{ \cal O}({}^3S_{1,8})|
\Upsilon \right\rangle$. According to the velocity scaling rule and
taking $v^2=0.08$, we might naively have
\begin{eqnarray}\left\langle\Upsilon|{ \cal O}({}^3S_{1,8})| \Upsilon
\right\rangle &\approx& \frac{v^4}{2N_c} \left\langle\Upsilon|{ \cal
O}({}^3S_{1,1})| \Upsilon \right\rangle =4.1 \times 10^{-3}~ {\rm
GeV^3}.
\end{eqnarray}
Using Eq.~(\ref{eq:bbccOCTNLO}), we would get
\begin{eqnarray}
B_{NLO}[\Upsilon({}^3S_{1,8}) \to c \bar c+X]&=&21\%.
\end{eqnarray}
For the light quark $q=u,d,s$, we have
\begin{eqnarray}
&&B_{NLO}[\Upsilon({}^3S_{1,8}) \to q \bar
q+X]=B_{NLO}[\Upsilon({}^3S_{1,8}) \to c \bar c+X] \times\left(1+
{\cal O}\left(r^2\right)\right).
\end{eqnarray}
So $\Upsilon$ decay through $b \bar b( {}^3S_{1,8})$ would have a
very large branching ratio, say about $~80\%$. Apparently, the color
-octet matrix element estimated in this naive way from the velocity
scaling rule is greatly overestimated, even by an order of
magnitude.

Another approach to determine the matrix element is the lattice QCD
calculations. The lattice calculation in Ref.~\cite{Bodwin:2005gg}
gives
\begin{eqnarray}
\left\langle\Upsilon|{ \cal O}({}^3S_{1,8})| \Upsilon \right\rangle
&\approx& 8.1 \times10 ^{-5} \left\langle\Upsilon|{ \cal
O}({}^3S_{1,1})| \Upsilon \right\rangle =3.1 \times 10^{-4}~ {\rm GeV^3}.
\end{eqnarray}
If we set the soft cut $\delta_s =0.15$, then the next-to-leading
order result is
\begin{eqnarray}
B_{NLO}[\Upsilon({}^3S_{1,8}) \to c \bar c]&=&1.3\%.
\end{eqnarray}
If we set the soft cut $\delta_s =0.10$ and $\delta_s =0.20$, the
branching ratio is $1.1\%$ and $1.4\%$ respectively.

From the above numerical results and Eq.(\ref{eq:bbccOCTNLO}) we see
that since the short distance coefficient for the color-octet
contribution to the $\Upsilon\to c\bar c$ decay is large,  this
process is sensitive to the value of the color-octet matrix element,
and may therefore serve as a useful test ground of the color-octet
mechanism.

Moreover, the next-to-leading order QCD correction in the
color-singlet piece is much stronger than that in the color-octet
piece, and the color-singlet contribution shows a strong sensitivity
to the soft cut $\delta_s$, whereas the color-octet result does not.
These differences between the color-singlet and color-octet
contributions will also be significant in clarifying the issue about
the color-octet mechanism. Once the experiment can measure the
branching ratio and energy distribution of the charm quark jet in
the $\Upsilon$ decay, the result can be used to test the color-octet
mechanism or give a strong constraint on the color-octet matrix
elements.

\section{Summary and discussion}
\label{sec:summ&dis} We calculate the decay rate of bottomonium to
two-charm quark jets $\Upsilon \to c \bar c$ at the tree level and
one-loop level including color-singlet and color-octet $b \bar b$
annihilations. We find that the short distance coefficient of the
color-octet piece is much larger than the color-singlet piece, and
that the QCD correction will change the endpoint behavior of the
charm quark jet. The color-singlet piece is strongly affected  by
the one-loop QCD correction. In contrast, the QCD correction to the
color-octet piece is weak. Once the experiment can measure the
branching ratio and energy distribution of the two-charm quark jets
in the $\Upsilon$ decay, the result can be used to test the color
octet mechanism or give a strong constraint on the color-octet
matrix elements.
\newline

After our work was completed \cite{zhang:2007phd}, a paper appeared
\cite{Kang:2007uv}, in which Kang, Kim, Lee, and Yu calculated the
inclusive charm production in $\Upsilon(nS)$ decay. They focused on
the inclusive charm production of the color-singlet piece at leading
order in the strong coupling constant $\alpha_s$. We focused on the
$c \bar c$ final state and the color-octet mechanism. The $c \bar c$
final state is essentially the two charm-jet process. And we have
calculated the next-to-leading order QCD corrections in both color
singlet and color octet pieces. Our leading order result of
$\Upsilon \to \gamma^* \to c \bar c$ is consistent with their result
\cite{Kang:2007uv}.

\begin{acknowledgments}
We thank C. Meng for useful discussions. This work was supported in
part by the National Natural Science Foundation of China (No
10421503, No 10675003), and also by China Postdoctoral Science
Foundation (No 20070420011).

\end{acknowledgments}

\vspace*{1cm}

\appendix

\noindent {\large \bf APPENDICES}

\section{The scalar functions}\label{APP1}
The scalar functions that appear in the virtual corrections are
listed in this Appendix. There are UV, IR and Coulomb singularities
in the scalar functions. The UV and IR singularities are
regularized with $D=4-2\epsilon$ space-time dimension. 
The exchange of longitudinal gluon between massive quarks in Vertex
N1 in Fig.\ref{bbccSLLoop} and Fig.\ref{bbccOCTLOOP} leads to a
Coulomb singularity $\sim \pi^2/v$, where $v=\sqrt{-(p_{b}-p_{\bar
b})^2}/m_b$ is the relative velocity between $b$ and $\bar{b}$ in
the meson rest frame ($v=|\overrightarrow{p_{b}}-
\overrightarrow{p_{\bar b}}|/m_b$).   The Coulomb singularities
should be canceled by that in the matrix elements (see, e.g.,
\cite{Zhang:2005ch,Zhang:2006ay}).

Since the imaginary part of the integrals will disappear in the
final result, only the real parts are given. The external particles
are taken to be on-mass-shell, $p_b^2=p_{\bar b}^2=m_b^2$,
$p_c^2=p_{\bar c}^2=m_c^2$, $p_b \cdot p_c=p_b \cdot p_{\bar
c}=m_b^2$, and $p_c \cdot p_{\bar c}=2 m_b^2-m_c^2$.

The scalar one-point function is defined as
\begin{eqnarray} A_0(m^2) = \mu^{4-D}\;\int\frac{\mbox{d}^Dq}{(2\pi)^D}\;
\frac{1}{q^2-m^2} =
iC_{\epsilon}(m)\;m^2\,\left[\,\frac{1}{\epsilon}+1\,\right]
\end{eqnarray} where \begin{eqnarray}\label{EQ_CEPS} C_{\epsilon}(m) =
\frac{1}{16\pi^2} \;\; e^{-\epsilon(\gamma_E-\ln 4\pi)}
           \left(\frac{\mu^2}{m^2}\right)^{\epsilon}
\end{eqnarray} and $D=4-2\epsilon$.

The scalar two-point function is defined as
\begin{eqnarray} B_0(p,m_0,m_1) =
\mu^{4-D}\;\int\frac{\mbox{d}^Dq}{(2\pi)^D}\;
\frac{1}{[q^2-m_0^2]\,[(q+p)^2-m_1^2]} \quad
\end{eqnarray}
Four different types of two-point functions appear in the
calculation of the virtual corrections:
\begin{eqnarray} B_0(p_b,0,m_b)&=&
B_0(2p_b,m_b,m_b) =
      iC_{\epsilon}(m_b)\;\left[\,\frac{1}{\epsilon}+2\,\right] \\[1mm]
B_0(p_c,0,m_c)&=& B_0(p_{\bar c},0,m_c) =
      iC_{\epsilon}(m_c)\;\left[\,\frac{1}{\epsilon}+2\,\right] \\[1mm]
 B_0(2p_b,m_c,m_c) &=&
iC_{\epsilon}(m_c)\;\left[\,
       \frac{1}{\epsilon}+2+\beta\ln\left(\frac{1-\beta}
       {1+\beta}\right)
       \right] \\[1mm]
 B_0(p_c-p_b,m_c,m_b) &=&
iC_{\epsilon}(m_c)\;\left[\,
       \frac{1}{\epsilon}+2+\frac{1}{\beta}\ln\left(\frac{1-\beta}
       {1+\beta}\right)
       \right] \\[1mm]
 B_0(2p_b,0,0) &=& iC_{\epsilon}(m_c)\;\left[\,
       \frac{1}{\epsilon}-\ln\left(\frac{4 m_b^2}{m_c^2}\right)+2
       \right].\end{eqnarray}
Here and below we will use the shorthand notation
$\beta=\sqrt{1-r^2}=\sqrt{1-m_c^2/m_b^2}$.

The scalar three-point function is defined as
\begin{eqnarray} \lefteqn{\hspace*{-1cm}
C_0(p_1,p_2,m_0,m_1,m_2) = }\nonumber\\[1mm]
& & \mu^{4-D}\;\int\frac{\mbox{d}^Dq}{(2\pi)^D}\;
\frac{1}{[q^2-m_0^2]\,[(q+p_1)^2-m_1^2]\,[(q+p_2)^2-m_2^2]} \quad .
\end{eqnarray}
The following types of three-point functions appear in the virtual
corrections: 

\begin{eqnarray}\label{eq:bas3pfc}
C_0(p_c,-p_{\bar c},0,m_c,m_c)&=& \frac{i
C_\epsilon(m_c)}{4m_b^2\beta}\left[\frac{1}{\epsilon}\ln
x_{\beta}-2\ln x_{\beta}\ln(1-x_{\beta})-2
{{\mbox{Li}}}_2(x_{\beta})\right. \nonumber \\
&&\hspace{0.5cm}\left.+\frac{1}{2}\ln^2x_{\beta}-4\zeta(2)\right]
\\C_0(p_c,p_{b},0,m_c,m_b)&=& \frac{i
C_\epsilon(\sqrt{m_cm_b})}{4m_c m_b
\chi/(\chi^2-1)}\left[\frac{1}{\epsilon}\ln \left(\frac{1-\chi
}{\chi +1}\right) +\frac{1}{2} \ln ^2\left(\frac{1-\chi
   }{\chi +1}\right) -\frac{\ln ^2r}{2}\right.\nonumber \\
   &&\hspace{0.5cm}\left.  -2 \ln \left(\frac{4 \chi }{(\chi
   +1)^2}\right) \ln \left(\frac{1-\chi }{\chi
   +1}\right)-\text{Li}_2\left(\frac{(\chi -1)^2}{(\chi
   +1)^2}\right)\right.\nonumber \\
   &&\hspace{0.5cm}\left.-\text{Li}_2\left(1+\frac{r (\chi -1)
   }{\chi +1}\right)-\text{Li}_2\left(1+\frac{\chi
   -1}{r (\chi +1)}\right)+\frac{\pi ^2}{6} \right]
\\
C_0(p_c,-p_{\bar c},m_c,0,0)&=&i  \frac{1}{4(4
\pi)^2m_b^2\beta}\left[2 {{\mbox{Li}}}_2(-x_{\beta})+
\frac{1}{2}\ln^2x_{\beta}+\zeta(2)\right]\\
C_0(p_b,-p_{b},m_b,0,0)&=&i  \frac{\ln 2}{(4 \pi)^2 m_b^2},
\end{eqnarray}
where $x_{\beta}=(1-\beta)/(1+\beta)$ ,  $\zeta(2) = \pi^2/6$,
$r=m_c/m_b$, and $\chi=\sqrt{(1-r)/(1+r)}$. There is another scalar
three-point function that is IR and Coulomb divergent,
\begin{eqnarray}
C_0(p_b,-p_{\bar b},0,m_b,m_b)&=&-i  \frac{C_\epsilon(m_b)}{
   2m_b^2}\left[\,
\frac{1}{\epsilon} + \frac{\pi^2}{v} -2   + {\cal O}(\epsilon)
\right]
\end{eqnarray}
where $v=\sqrt{-(p_{b}-p_{\bar b})^2}/m_b$. In the meson rest frame,
we have $v=|\overrightarrow{p_{b}}- \overrightarrow{p_{\bar
b}}|/m_b$.

The scalar four-point function is defined by
\begin{eqnarray}
&& \hspace{-1.3cm}D_0(p_1,p_2,p_3,m_0,m_1,m_2,m_3) = \nonumber\\[1mm]
&& \hspace{-1.0cm} \mu^{4-D}\;\int\frac{\mbox{d}^Dq}{(2\pi)^D}\;
\frac{1}{[q^2-m_0^2]\,[(q+p_1)^2-m_1^2]\,[(q+p_2)^2-m_2^2]
\,[(q+p_3)^2-m_3^2]} \quad . \end{eqnarray} There are three
different types of four-point functions: \begin{eqnarray} &&
D_0(p_b,p_b-p_c,-p_b,m_b,0,m_c,0) \nonumber \\&=&\frac{i
C_\epsilon(\sqrt{m_cm_b})}{8m_b^4} \left\{\frac{(\chi^2-1)}{r
\chi}\left[\frac{1}{\epsilon}\ln \left(\frac{1-\chi }{\chi
+1}\right) +\frac{1}{2} \ln ^2\left(\frac{1-\chi
   }{\chi +1}\right) -\frac{\ln ^2r}{2}\right.\right.\nonumber \\
   &&\left.  -2 \ln \left(\frac{4 \chi }{(\chi
   +1)^2}\right) \ln \left(\frac{1-\chi }{\chi
   +1}\right)-\text{Li}_2\left(\frac{(\chi -1)^2}{(\chi
   +1)^2}\right)-\text{Li}_2\left(1+\frac{r (\chi -1)
   }{\chi +1}\right)\right.\nonumber \\
   &&\left.\left.-\text{Li}_2\left(1+\frac{\chi
   -1}{r (\chi +1)}\right)+\frac{\pi ^2}{6} \right] -
   \frac{1}{\beta}\left[2 {{\mbox{Li}}}_2(-x_{\beta})+
\frac{1}{2}\ln^2x_{\beta}+\zeta(2)\right]\right\},
\end{eqnarray}

The IR and Coulomb singularities can be regularized by the gluon
mass $m_g$. The relation between the gluon mass $m_g$ regularization
and the dimensional regularization for IR singularity is
\begin{eqnarray}\label{eq:IRCor}
\ln \left(\frac {\lambda^2}{m^2}\right) \Longleftrightarrow
\frac{1}{\epsilon}-\gamma_\mathrm{E}+ \ln
\frac{4\pi\mu^2}{m^2}
\end{eqnarray}
And the relation between different regularization schemes for the
Coulomb singularity is
\begin{eqnarray}\label{eq:coulombCor}
 \frac{2 \pi m}{\lambda} \Longleftrightarrow
 \frac{\pi^2}{v}
\end{eqnarray}
Eq.(\ref{eq:IRCor}) and Eq.(\ref{eq:coulombCor}) are consistent with
Ref.\cite{Kramer:1995nb}.

\section{\label{appa} Real corrections and the three-body phase space }
 For the real corrections, the
process $\Upsilon(2p_b)\to  c(p_c)+\bar c(p_{\bar c})+g(k)$ is a
three-body decay process. Similar to the method in
Ref.~\cite{Hahn:1998yk}, we can write down the Lorentz-invariant
phase space
\begin{eqnarray}
\rm d PS_3(2p_b;k,p_c,p_{\bar c})&=& \frac{\rm d^3 k}{(2\pi)^3 2
k^0} \frac{\rm d^3 p_c}{(2\pi)^3 2 p_c^0}\frac{\rm d^3 p_{\bar
c}}{(2\pi)^3 2 p_{\bar c}^0} (2 \pi)^4 \delta ^4 (2p_b-k-p_c-p_{\bar
c}).
 \end{eqnarray}
Introduce the identities
\begin{eqnarray}\label{identities}
\frac{\rm d^3 p_i}{2 p_i^0}=\rm d^4 p_i \delta(p_i^2-m_i^2)={\rm
\frac{|\overrightarrow{p_i}|^2 \  d|\overrightarrow{p_i}|\  d
\Omega_i}{2 p_i^0}}={\rm \frac {|\overrightarrow{p_i}| \ dp_i^0 \ d
\Omega_i}{2}},
 \end{eqnarray}
where $m_i$ is the mass of particle $i$, and $d \Omega_i$ is the
direction of particle $i$ in the $3$ dimension space. Then we can
rewrite $\rm d PS_3(2p_b;k,p_c,p_{\bar c})$:
 \begin{eqnarray}
\rm d PS_3&=&\frac{|\overrightarrow{k}| |\overrightarrow{p_c}|}{4 (2
\pi)^5}\  d k^0 \  d \Omega_g \  d p_c^0 \  d \Omega_c  \ d^4
p_{\bar c}\  \delta (p_{\bar c}^2-m_{\bar c}^2)\ \delta ^4
(2p_b-k-p_c-p_{\bar c}) \nonumber
\\ &=&\frac{|\overrightarrow{k}| |\overrightarrow{p_c}|}{4
(2 \pi)^5}\  d k^0\  d \Omega_g \ d p_c^0 \ d \Omega_c\  \delta[
(2p_b-k-p_c)^2-m_{\bar c}^2]
\end{eqnarray}
We define the momenta in the rest frame of the  $\Upsilon$,
\begin{eqnarray}
2p_b&=&(2m_b ,0,0,0)  \hspace{1.5 cm}
p_c=(p_c^0,|\overrightarrow{p_c}|
\sin \theta, 0, |\overrightarrow{p_c}| \cos \theta) \nonumber \\
k&=&(k^0,0,0,|\overrightarrow{k}|)
 \hspace{1.2 cm} p_{\bar
c}=(p_{\bar c}^0,-|\overrightarrow{p_c}| \sin \theta, 0,
-|\overrightarrow{k}|-|\overrightarrow{p_c}| \cos \theta),
 \end{eqnarray}
 where $\theta$ is the angular between $g $ and $c$ , and
  $|\overrightarrow{p_i}|=\sqrt{(p_i^0)^2-m_i^2}$.
Then $d \Omega_g$ gives a factor $4 \pi$, and $d \Omega_c= d \cos
\theta d \phi$ and $ d \phi$ gives a factor $2 \pi$. So we have
 \begin{eqnarray}
\rm d PS_3&=&\frac{|\overrightarrow{k}| |\overrightarrow{p_c}|}{2 (2
\pi)^3}\  d k^0 \  d p_c^0 \  d \cos \theta \ \delta[
(2p_b-k-p_c)^2-m_{\bar c}^2].
\end{eqnarray}
Then we use the $\delta$ function to remove $\theta $ in the
integral with
\begin{eqnarray}
(2p_b-k-p_c)^2-m_{\bar c}^2&=&(\sqrt s
-k^0-p_c^0)^2-(|\overrightarrow{k}|^2+ |\overrightarrow{p_c}|^2 + 2
|\overrightarrow{k}| |\overrightarrow{p_c}| \cos \theta)- m_{\bar
c}^2 \nonumber \\ &\equiv&f(\cos \theta)
\end{eqnarray} and
\begin{eqnarray}
\left | \frac{d f(\cos \theta)}{d \cos \theta}\right |&=&2
|\overrightarrow{k}| |\overrightarrow{p_c}|,
\end{eqnarray}
and  get
\begin{eqnarray}
\cos \theta&=& \frac{(\sqrt s -k^0-p_c^0)^2 -|\overrightarrow{k}|^2-
|\overrightarrow{p_c}|^2 - m_{\bar c}^2}{2 |\overrightarrow{k}|
|\overrightarrow{p_c}|},\end{eqnarray} and
\begin{eqnarray}\rm d PS_3&=&\frac{1}{4 (2 \pi)^3}\  d k^0 \  d
p_c^0 \ .
\end{eqnarray}
To determine the limits of integration, we employ the restriction of
$|\cos \theta|\leq 1$ and $p_i^0 \geq m_i^0$, then we get
\begin{eqnarray}
\left ( k^0\right)^{min}&=& m_g, \nonumber \\ \left (
k^0\right)^{max}&=&m_b -\frac{(m_c +m_{\bar c})^2- m_g^2}{4m_b},
\end{eqnarray}
and
\begin{eqnarray} \label{eq:3bodyPS}
&&\left ( p_c^0 \right)^{max,min} =\frac{1}{2 \tau}
\left[\sigma(\tau+m_+m_-)\pm
|\overrightarrow{k}|\sqrt{(\tau-m_+^2)(\tau-m_-^2)}\ \right]
\nonumber \\
&&\hspace{0.5 cm} \sigma= \sqrt s-k^0, \hspace{1 cm} \tau=
\sigma^2-|\overrightarrow{k}|^2, \hspace{1 cm}  m_\pm =m_c \pm
m_{\bar c}.
\end{eqnarray}
Here we keep the gluon mass $m_g$ for massive gluon regularization.
There is a soft divergence in the real corrections, so we should
introduce a soft cut $E_s$ for the gluon. Then the phase space is
divided into two regions:
\begin{eqnarray}\rm d PS_3&=&\rm  d PS_3^{Soft}\left|_{k^0<E_s}
+\rm  d PS_3^{Hard}\right|_{k^0>E_s}.
\end{eqnarray}
The hard region can be integrated in four dimension or with massless
gluon. And the phase space in the soft region is
\begin{eqnarray}
\left.{\rm  d PS_3^{Soft}}\right|_{k^0<E_s} &=& {\rm  d PS_2}\int
{\rm d \Omega_g^{D-1}} \int^{E_s} \frac{|\overrightarrow
k|^{D-3}}{2(2\pi)^{D-1}}{\rm d}k^0.
\end{eqnarray}
The decay amplitude of the color singlet process can be written as
\begin{eqnarray}
&&\hspace{-2cm}\left.{\cal A}^{Real}\Big(b\bar{b}({}^{3}S_{1,
1}(2p_b)\rightarrow c(p_c)+ \bar{c}(p_{\bar c})+g(k)\Big)
\right|_{k^0<E_s}=\nonumber
\\&&g_s\mu^\epsilon \varepsilon_{*\mu}^a (k)
 {\cal A}^{Born}\Big(b\bar{b}({}^{3}S_{1, 1}(2p_b)\rightarrow c(p_c)+
\bar{c}(p_{\bar c})\Big)\otimes T^a \left( \frac{p_c^\mu}{p_c \cdot
k} -\frac{p_{\bar c}^\mu}{p_{\bar c} \cdot k}\right)
\end{eqnarray}
and
\begin{eqnarray}
\left|\left.{\cal A}^{Real}\Big({}^{3}S_{1, 1}\Big)
\right|_{k^0<E_s}\right|^2=\left|{\cal A}^{Born}\Big({}^{3}S_{1,
1}\Big)\right|^2 g_s^2 \mu^{2\epsilon} \frac{4}{3} \left( {\cal
I}_{cc}-2{\cal I}_{c \bar c} +{\cal I}_{\bar  c \bar c}\right),
\end{eqnarray}
where
\begin{eqnarray}
{\cal I}_{ij}=-\frac{p_i \cdot p_j}{p_i \cdot k p_j \cdot k}.
\end{eqnarray}
The decay amplitude of the  color octet process can be written as
\begin{eqnarray}
&&\hspace{-2cm}\left.{\cal A}^{Real}\Big(b\bar{b}({}^{3}S_{1,
8}(2p_b)\rightarrow c(p_c)+ \bar{c}(p_{\bar c})+g(k)\Big)
\right|_{k^0<E_s}=\nonumber
\\&&g_s\mu^\epsilon \varepsilon_{*\mu}^a(k)\left( \frac{p_c^\mu}{p_c
\cdot k} T^a \otimes {\cal A}^{Born} -{\cal A}^{Born}\otimes  T^a
\frac{p_{\bar c}^\mu}{p_{\bar c} \cdot k}-{\cal A}^{Born}\frac{i
f^{ab_\Upsilon c_{c \bar c}}}{2} \frac{p_b^\mu}{p_b \cdot k}\right)
\end{eqnarray}
and

\begin{eqnarray}
&&\left|\left.{\cal A}^{Real}\Big({}^{3}S_{1, 8}\Big)
\right|_{k^0<E_s}\right|^2=\nonumber \\
&&\hspace{1cm}\left|{\cal A}^{Born}\Big({}^{3}S_{1, 8}\Big)\right|^2
g_s^2 \mu^{2\epsilon} \left( \frac{4}{3} {\cal
I}_{cc}+\frac{1}{3}{\cal I}_{c \bar c} +\frac{4}{3} {\cal I}_{\bar c
\bar c}+3 {\cal I}_{bb}-3 {\cal I}_{b c }-3 {\cal I}_{b\bar c
}\right).
\end{eqnarray}
The integration of ${\cal I}_{ij}$ in the soft region with
dimensional regularization can be found in Ref.\cite{Harris:2001sx}.
and with massive gluon can be found in Ref.\cite{'tHooft:1978xw}

\section{the total decay width}
The decay width at NLO in $\alpha_s$ is
\begin{eqnarray}
\Gamma_{NLO}=\Gamma_{LO}+\Gamma_{Virtual}+\Gamma_{Real}.
\end{eqnarray}
The LO decay width of the color singlet piece has been given in
Eq.(\ref{eq:LOSOcc}). The D-dimension LO decay width is
\begin{eqnarray}
\Gamma_{LO}[\Upsilon({}^3S_{1,1}) \to c \bar c]=\frac{4\pi \alpha^2
\beta  \left(D-2
   +r^2\right)   }{81 m_b^2}\frac{\sqrt\pi
   }{2\Gamma[\frac 3 2
   -\epsilon]}\left(\frac{4\pi\mu^2}{\beta^2m_b^2}\right)^\epsilon
   \left\langle\Upsilon|{ \cal O}({}^3S_{1,1})| \Upsilon
   \right\rangle,
\end{eqnarray}
where $r=m_c/m_b$ and $\beta=\sqrt{1-r^2}$.

The $\Gamma_{Virtual}$  of the color singlet piece can be written as
\begin{eqnarray}
&&\hspace{-0.5cm}\frac{\Gamma_{Virtual}({}^3S_{1,1})}{\Gamma_{LO}({}^3S_{1,1})}=\frac{
\alpha _s}{ \pi }\frac{r^2+2}{r^2+D-2}\Bigg\{
-4\left(\frac{1}{\epsilon }-\gamma_E + \ln\left(\frac{4 \pi  \mu
^2}{m_b m_c}\right)\right) +\frac{16 m_b^2}{3 \left(2
m_b^2+m_c^2\right)}
   -\frac{52}{9}+ \nonumber \\ &&
   \frac{32 i \pi ^2}{9 \left(2 m_b^2+m_c^2\right)}\Big[
   {A_0}\left(m_b\right)
   \left(4-\frac{m_c^2}{m_b^2}\right)-3
   {A_0}\left(m_c\right) -2 \Big(3
   {B_0}\left(p_c,0,m_c\right) \left(4
   m_b^2+m_c^2\right)+  \nonumber \\ && \left.
   {B_0}\left(2 p_b,m_b,m_b\right) \left(2
   m_b^2+m_c^2\right)-3 {B_0}\left(2p_b,m_c,m_c\right) \left(3 m_b^2+m_c^2\right)
   +3 {B_0}\left(p_b,0,m_b\right)
   m_b^2\Big)\right. \nonumber \\ && +12 \Big((D-2) m_b^2+m_c^2\Big)
   \Big({C_0}\left(p_b,-p_{\bar b},0,m_b,m_b\right) m_b^2 +\nonumber \\ && {C_0}\left(p_c,-p_{\bar c},0,m_c,m_c\right) \left(2
   m_b^2-m_c^2\right)\Big)\Big]\Bigg\}.
%
%
\end{eqnarray}

For the real corrections, we should introduce the Mandelstam
variables
\begin{eqnarray}
s_{23}&=&(p_c+p_{\bar c})^2 \nonumber \\
s_{34}&=&(p_{\bar c}+k)^2.
\end{eqnarray}
In the rest frame of $\Upsilon$,
\begin{eqnarray}
s_{23}&=&4m_b^2-4m_bk^0 \nonumber \\
s_{34}&=&4m_b^2+m_c^2-4m_b p_c^0.
\end{eqnarray}
The contribution of real corrections for the color singlet piece is
\begin{eqnarray}
&&\hspace{-0.5cm}\frac{{\rm d}\Gamma_{Real}({}^3S_{1,1})}{{\rm d}k^0
{\rm d} p_c^0}= -\frac{16 \left\langle\Upsilon|{ \cal
O}({}^3S_{1,1})| \Upsilon
   \right\rangle \alpha ^2 \alpha _s}{243  m_b^4
   \left(m_c^2-s_{34}\right){}^2 \left(-4
   m_b^2-m_c^2+s_{23}+s_{34}\right){}^2} \times\nonumber \\
   &&\Bigg\{2 m_c^8-8 s_{34}
   m_c^6+\left(3 s_{23}^2+4 s_{34} s_{23}+12 s_{34}^2\right)
   m_c^4-\left(s_{23}^3+2 s_{34} s_{23}^2+8 s_{34}^2 s_{23}+8
   s_{34}^3\right) m_c^2  + \nonumber \\  && \left.
   64 m_b^6 \left(3 m_c^2-s_{34}\right)+s_{34} \left(s_{23}+s_{34}\right)
   \left(s_{23}^2+2 s_{34} s_{23}+2 s_{34}^2\right) +\right. \nonumber \\  && \left.16 m_b^4
   \Big[7 m_c^4-\left(5 s_{23}+6 s_{34}\right) m_c^2+s_{34}
   \left(s_{23}+3 s_{34}\right)\Big]+\right. \nonumber \\  &&
   4 m_b^2 \Big[4 m_c^6-4
   \left(2 s_{23}+3 s_{34}\right) m_c^4+\left(3 s_{23}^2+4 s_{34}
   s_{23}+12 s_{34}^2\right) m_c^2-s_{34} \left(s_{23}+2
   s_{34}\right){}^2\Big]\Bigg\}.
\end{eqnarray}

 The NLO decay width of color singlet
piece is
\begin{eqnarray} &&\hspace{-1cm}\Gamma_{NLO}[({}^3S_{1,1})]
  = \frac{8\pi \alpha^2
  }{81 m_b^2}
   \left\langle\Upsilon|{ \cal O}({}^3S_{1,1})| \Upsilon
   \right\rangle \Bigg\{ \sqrt{1-r^2} \left(1
   +\frac{r^2}{2}\right) \left( 1-\frac{16\alpha_s}{4\pi} C_F\right)+\nonumber \\&&
   \frac{\alpha_s}{4\pi} C_F
      \Bigg[(32-8r^4){\rm Li}_2{(x_{\beta})} +
      (16-4r^4)\Big({\rm Li}_2{(-x_{\beta})}+\ln(x_{\beta})\ln(1-x_{\beta})\Big)
        \nonumber \\
& &+
(2+r^2)\sqrt{1-r^2}\Big(6\ln(x_{\beta})-8\ln(1-x_{\beta})-4\ln(1+x_{\beta})\Big)
  +
(3+\frac{9r^2}{2})\sqrt{1-r^2}\nonumber  \\
& & + (-12+2r^2+\frac{7r^4}{4})\ln(x_{\beta})+
(8-2r^4)\ln(x_{\beta})\ln(1+x_{\beta}) \Bigg] \Bigg\}\, ,
\label{eqn:anal_massive}
\end{eqnarray}
where $\beta=\sqrt{1-r^2}$ and
$x_{\beta}=(1-\beta)/(1+\beta)=(1-\sqrt{1-r^2})/(1+\sqrt{1-r^2})$.
The $-\frac{16\alpha_s}{4\pi} C_F$ term is due to  the QCD
correction to $\Upsilon[b \bar b({}^3S_{1,1})] \to \gamma^*$, while
the $\frac{\alpha_s}{4\pi} C_F[\cdots]$ term is due to the QCD
correction to $\gamma^* \to c \bar c$. This is the same as the known
next-to-leading order result of $e^+e^- \to \gamma^* \to c \bar
c$~\cite{Jersak:1981sp,Ravindran:1998qz}.

The LO decay width of the color octet piece has been given in
Eq.(\ref{eq:LOOCTcc}). The D-dimension LO decay width is
\begin{eqnarray}
\Gamma_{LO}[\Upsilon({}^3S_{1,8}) \to c \bar c]= \frac{\alpha_s^2
\beta \left(D-2
  +r^2\right)  \pi }{6 m_b^2}\frac{\sqrt\pi
   }{2\Gamma[\frac 3 2
   -\epsilon]}\left(\frac{4\pi\mu^2}{\beta^2m_b^2}\right)^\epsilon
   \left\langle\Upsilon|{ \cal O}({}^3S_{1,8})| \Upsilon
   \right\rangle.
\end{eqnarray}

The $\Gamma_{Virtual}$  of the color singlet piece can be written as
\begin{eqnarray}
&&\hspace{-0.5cm}\frac{\Gamma_{Virtual}({}^3S_{1,8})}{\Gamma_{LO}({}^3S_{1,8})}=\frac{
\alpha _s(\mu)}{9 \pi }\frac{r^2+2}{r^2+D-2}\Bigg\{-\frac{147}{2
}\left(\frac{1}{\epsilon }-\gamma_E + \ln\left(4 \pi \right)
\right)-18 \log \left(\frac{\mu ^4}{m_b^2 m_c^2}\right) +\nonumber
\\ && \frac{202 m_b^2-193 m_c^2}{4 \left(2 m_b^2+m_c^2\right)}
+\frac{2 i \pi ^2}{m_b^2 \left(2
m_b^2+m_c^2\right)}\Big[-{A_0}\left(m_b\right) \left(35 m_b^2-2
m_c^2\right)-12
   {A_0}\left(m_c\right) \left(6 m_b^2+m_c^2\right)-\nonumber
\\ &&15 {B_0}\left(p_b,0,m_b\right) \left(7 m_b^4+12 m_c^2
   m_b^2\right)-90 {B_0}\left(2p_b,0,0\right) m_b^2 \left(2
   m_b^2+m_c^2\right)+ \nonumber
\\ &&4 {B_0}\left(2p_b,m_b,m_b\right) m_b^2
   \left(2 m_b^2+m_c^2\right)+12 {B_0}\left(2p_b,m_c,m_c\right)
   \left(m_b^4+3 m_c^2 m_b^2+m_c^4\right)- \nonumber
\\ &&96
   {B_0}\left(p_c,0,m_c\right) m_b^2 \left(4
   m_b^2+m_c^2\right)-108
   {B_0}\left(p_b-p_c,m_b,m_c\right) m_b^4
- \nonumber
\\&&324 {C_0}\left(p_b,-p_b,m_b,0,0\right) m_b^4
   \left(2 m_b^2+m_c^2\right)-216 {C_0}\left(p_c,-p_{\bar c},m_c,0,0\right) m_b^4 \left(2
   m_b^2+m_c^2\right)- \nonumber
\\&&24 {C_0}\left(p_b,-p_{\bar b},0,m_b,m_b\right) m_b^4 \left((D-2) m_b^2+m_c^2\right)- \nonumber
\\&&216
   {C_0}\left(p_c,p_b,0,m_c,m_b\right) m_b^4
   \left((D-2) m_b^2+m_c^2\right)- \nonumber
\\&&24 {C_0}\left(p_c,-p_{\bar c},0,m_c,m_c\right) m_b^2 \left(2
   m_b^2-m_c^2\right) \left((D-2) m_b^2+m_c^2\right)
- \nonumber
\\ &&432 D_0(p_b,p_b-p_c,-p_b,m_b,0,m_c,0) m_b^6 \big((D-2)
   m_b^2+m_c^2\big)\Big]\Bigg\}.
\end{eqnarray}

The contribution of real corrections for the color octet piece is
\begin{eqnarray}
&&\hspace{-0.5cm}\frac{{\rm d}\Gamma_{Real}({}^3S_{1,8})}{{\rm d}k^0
{\rm d} p_c^0}=-\frac{\left\langle\Upsilon|{ \cal O}({}^3S_{1,8})|
\Upsilon   \right\rangle   \alpha
   _s^3}{18 m_b^4 \left(s_{23}-4 m_b^2\right){}^2
   \left(m_c^2-s_{34}\right){}^2 \left(-4
   m_b^2-m_c^2+s_{23}+s_{34}\right){}^2}\times\nonumber
   \\&&\Big[64 m_b^4+4 \left(9 m_c^2-8
s_{23}-9
   s_{34}\right) m_b^2+9 m_c^4+4 s_{23}^2+9 s_{34}^2+9 s_{23}
   s_{34}-9 m_c^2 \left(s_{23}+2 s_{34}\right)\Big]\times\nonumber
   \\&& \Bigg\{2
   m_c^8-8 s_{34} m_c^6+\left(3 s_{23}^2+4 s_{34} s_{23}+12
   s_{34}^2\right) m_c^4-\left(s_{23}^3+2 s_{34} s_{23}^2+8
   s_{34}^2 s_{23}+8 s_{34}^3\right) m_c^2+\nonumber \\  && \left.64 m_b^6 \left(3
   m_c^2-s_{34}\right)+s_{34} \left(s_{23}+s_{34}\right)
   \left(s_{23}^2+2 s_{34} s_{23}+2 s_{34}^2\right)+\right. \nonumber \\  && \left.16 m_b^4
   \Big[7 m_c^4-\left(5 s_{23}+6 s_{34}\right) m_c^2+s_{34}
   \left(s_{23}+3 s_{34}\right)\Big]+\right. \nonumber \\  &&
   4 m_b^2 \Big[4 m_c^6-4
   \left(2 s_{23}+3 s_{34}\right) m_c^4+\left(3 s_{23}^2+4
   s_{34} s_{23}+12 s_{34}^2\right) m_c^2-s_{34}
   \left(s_{23}+2 s_{34}\right){}^2\Big]\Bigg\}.
\end{eqnarray}

 The NLO decay width of color octet
piece is
\begin{eqnarray} &&\hspace{-1cm}\Gamma_{NLO}[({}^3S_{1,8})]
  =  \frac{\pi\alpha_s^2
   \left\langle\Upsilon|{ \cal O}({}^3S_{1,8})| \Upsilon
   \right\rangle}{6 m_b^2}\beta \left(3 -\beta^2\right)+\frac{\alpha_s^3
   \left\langle\Upsilon|{ \cal O}({}^3S_{1,8})| \Upsilon
   \right\rangle}{216 m_b^2} \Bigg\{-12 \beta ^5+\nonumber \\&&
   2 \pi ^2 \beta ^4 -478 \beta ^3+14 \pi ^2 \beta
   ^2+1614 \beta -60 \pi ^2 -150 \beta  \left(\beta ^2-3\right) \ln
   \left(\frac{\mu^2}{m_b^2}\right)+\nonumber \\&&
   6 \left(\beta ^2-3\right) \ln (2) \left[90 \beta +\left(\beta
   ^2-17\right) \ln (2)\right]+408 \beta  \left(\beta ^2-3\right) \ln (\beta
   )-\nonumber \\&& -6 \beta  \left(35 \beta ^2-87\right) \log (1-\beta ^2)
   -6 \left(\beta ^6-13 \beta ^4+4 \beta ^2+60\right)
   \log \left( \frac{1-\beta }{\beta +1}
   \right)+\nonumber \\&&6 \log (\beta +1) \Big[\log (2) \beta ^4-5
   \log (16) \beta ^2+\left(-2 \beta ^4+31
   \beta ^2-75\right) \log (\beta +1)+51 \log (2)\Big]+\nonumber \\&&
   6 \log (1-\beta ) \Big[-\log (8) \beta ^4+60 \log (2) \beta ^2+3 \left(\beta ^4-20
   \beta ^2+51\right) \log (\beta +1)-153 \log (2)\Big]-\nonumber \\&&12
   \left(\beta ^2-3\right) \Big[9 \text{Li}_2(\beta )+9
   \text{Li}_2\left(\frac{\beta }{\beta +1}\right)+\left(\beta ^2-17\right) \left(2
   \text{Li}_2\left(\frac{2 \beta }{\beta +1}\right)-\text{Li}_2\left(\frac{\beta
   +1}{2}\right)\right)\Big]-\nonumber \\&&
   12 \left(\beta ^4-20 \beta ^2+51\right) \log (\beta ) \log \left( \frac{1-\beta }{\beta +1}
   \right) \Bigg\}.
\end{eqnarray}


\end{document}